\documentclass[pra,aps,twocolumn,10pt,showpacs,groupedaddress,superscriptaddress,floatfix]{revtex4-1}
\usepackage{amsmath,amsfonts,amssymb,graphics,graphicx,epsfig,color,times}
\usepackage[utf8x]{inputenc}
\usepackage{bbm, dsfont} 
\usepackage{subfigure}
\usepackage{hyperref}
\usepackage[capitalize]{cleveref}
\usepackage{mathrsfs}
\usepackage{verbatim}
\begin{document}
\title{Spin-incoherent Luttinger liquid of one-dimensional SU(\texorpdfstring{$\boldsymbol{\kappa}$}{kappa}) fermions}
\author{H. H. Jen}
\affiliation{Institute of Physics, Academia Sinica, Taipei 11529, Taiwan}
\author{S.-K. Yip}
\affiliation{Institute of Physics, Academia Sinica, Taipei 11529, Taiwan}
\affiliation{Institute of Atomic and Molecular Sciences, Academia Sinica, Taipei 10617, Taiwan}

\date{\today}

\renewcommand{\r}{\mathbf{r}}
\newcommand{\f}{\mathbf{f}}

\def\be{\begin{align}}
\def\ee{\end{align}}
\def\bea{\begin{eqnarray}}
\def\eea{\end{eqnarray}}
\def\ba{\begin{array}}
\def\ea{\end{array}}
\def\bdm{\begin{displaymath}}
\def\edm{\end{displaymath}}
\def\red{\color{red}}

\begin{abstract}
We theoretically investigate one-dimensional (1D) SU($\kappa$) fermions in the regime of spin-incoherent Luttinger liquid. We specifically focus on the Tonks-Girardeau gas limit where its density is sufficiently low that effective repulsions between atoms become infinite. In such case, spin exchange energy of 1D SU($\kappa$) fermions vanishes and all spin configurations are degenerate, which automatically puts them into spin-incoherent regime. In this limit, we are able to express the single-particle density matrices in terms of those of anyons. This allows us to numerically simulate the number of particles up to $N=32$. We numerically calculate single-particle density matrices in two cases: (1) equal populations for each spin components (balanced) and (2) all $S_z$ manifolds included. In contrast to noninteracting multi-component fermions, the momentum distributions are broadened due to strong interactions. As $\kappa$ increases, the momentum distributions are less broadened for fixed $N$, while they are more broadened for fixed number of particle per spin component. We then compare numerically calculated high momentum tails with analytical predictions which are proportional to $1/p^4$, in good agreement. Thus, our theoretical study provides a comparison with the experiments of repulsive multicomponent alkaline-earth fermions with a tunable SU($\kappa$) spin-symmetry in the spin-incoherent regime. 
\end{abstract}
\maketitle
\section{Introduction}

Huge interests in one-dimensional (1D) quantum systems \cite{Giamarchi2004, Cazalilla2011, Guan2013} are renewed in the past decade due to the experimental achievements in trapping 1D ultracold bosonic \cite{Paredes2004, Kinoshita2004, Haller2009} and fermionic gases \cite{Moritz2005, Liao2010}. A fundamental distinction between identical bosons and fermions lies in quantum statistics where bosons tend to condense in the same quantum state below their characteristic temperature while fermions cannot occupy a single quantum state owing to Pauli exclusion principle. When spinless bosonic particles are tightly confined in a quasi-1D regime, they become strongly interacting and fermionized in so-called Tonks-Girardeau (TG) gas limit \cite{Tonks1936, Girardeau1960}. This regime can be reached in a dilute gas such that the effective atom-atom interactions become infinite. Recent studies focus on the ground states or their momentum distributions of spinless bosons \cite{Olshanii1998, Girardeau2001, Minguzzi2002, Olshanii2003, Papenbrock2003, Forrester2003, Xu2015}, quantum magnetism in spinful bosons \cite{Deuretzbacher2008, Deuretzbacher2014, Volosniev2014, Yang2015, Yang2016, Deuretzbacher2016} or Bose-Fermi mixtures \cite{Deuretzbacher2017, Decamp2017}, and broadened momentum distributions of spin-incoherent \cite{Cheianov2005, Fiete2007, Feiguin2010} spin-$1$ Bose Luttinger liquid \cite{Jen2016_spin1, Jen2017_spin1}. As for recent investigations of 1D spinful fermions \cite{Guan2013}, energy spectra and mapping of spin-chain model for SU($\kappa$) fermions have been investigated \cite{Laird2017}, exotic pairing phase of Fulde, Ferrell, Larkin, and Ovchinnikov (FFLO) state with finite center-of-mass momenta has been indirectly observed in a spin-$1/2$ Fermi gas \cite{Liao2010}, and two distinguishable fermions can fermionize like two noninteracting identical fermions by tuning interparticle interactions \cite{Zurn2012}. 

For spin-$F$ fermions, only $F+1/2$ s-wave scattering lengths $a_s^{j}$ with even $j=0,2,...,2F-1$ \cite{Yip1999} are required to describe interaction dynamics of the states with a total spin equal to $j$. In two-electron fermionic atoms, there is no hyperfine interaction between the electronic $J=0$ and nuclear spins $I>0$ in the ground state ($^1S_0$). Therefore, all scattering lengths become equal. Under this condition, SU($\kappa=2I+1$) spin symmetry can emerge \cite{Cazalilla2009, Gorshkov2010, Cazalilla2014} in alkaline-earth fermions $^{87}$Sr ($I=9/2$) \cite{Bonnes2012, Messio2012} or $^{173}$Yb ($I=5/2$) \cite{Taie2012} with tunable spins \cite{Pagano2014, Decamp2016} close to the regime of spin-incoherent Luttinger liquid (SILL) \cite{Fiete2007}. 

The SILL is a different universal class from conventional Luttinger liquid (LL) \cite{Giamarchi2004, Haldane1981}, which shows exponential decays of single-particle Green's functions other than power-law decays in the respective spin and charge sectors of LL. This spin-incoherent regime is first investigated in semiconductor quantum wire \cite{Cheianov2004, Fiete2004, Fiete2007}, which can be reached when the thermal energy of the system is higher than the energy splitting of different spin states while still low enough that collective charge excitations are suppressed. Other systems in SILL regime, for example, uniform two-component gas \cite{Cheianov2005}, $t$-$J$ models \cite{Feiguin2010, Penc1996, Penc1997}, and two-dimensional Hubbard models \cite{Hazzard2013, Zhou2014}, have also been investigated. 

Specifically, for 1D spinful Bose gas in TG gas limit, the spin-independent interaction becomes infinite such that spin Hamiltonian can be ignored and all spin configurations are degenerate. Under this condition, the spatial wave functions of the atoms take the Slater determinant form of noninteracting fermions, and TG spinful Bose gas automatically resides in the regime of SILL \cite{Jen2016_spin1, Jen2017_spin1}. Similarly for spinful fermions in TG gas limit, spin exchange energy of 1D SU($\kappa$) fermions vanishes and all spin configurations are degenerate, which again puts them in SILL regime. Away from the TG limit, the condition for achieving SILL however differs between bosons and fermions. For weakly interacting 1D Bose gas, one has SILL if the differences among $a^j_s$ for different $j$'s are sufficiently small \cite{Jen2016_spin1, Jen2017_spin1}. On the other hand, for noninteracting 1D Fermi gas, the sound and spin wave velocities are both equal to the Fermi velocity if populations of all the components are equal. Hence for a weakly interacting 1D Fermi gas, one does not have SILL even when the interaction is SU($\kappa$) symmetric.

In Refs.\cite{Jen2016_spin1, Jen2017_spin1}, we have investigated SILL 1D spin-$1$ Bose gas in TG gas limit. We find the evident broadening in either the total or spin-dependent momentum distributions in the sector of zero magnetization. We have also derived the $1/p^4$ asymptotic \cite{Minguzzi2002, Olshanii2003, Xu2015, Braaten2008-1, Braaten2008-2, Werner2009, Zhang2009}, and evaluated the coefficient, related to Tan's contact \cite{Tan2008, Barth2011}, up to $N=16$. Here we investigate spinful fermions with tunable SU($\kappa$) spin symmetry in SILL TG regime, and numerically calculate their momentum distributions without the restriction of zero magnetization. We also extend the particle number to $N=32$ by taking the advantage of anyonic statistics (or discrete Fourier transform) which significantly expedites the numerical calculations of single-particle density matrix. Thus, our study provides a comparison with the experiments of repulsive multicomponent alkaline-earth fermions with a tunable SU($\kappa$) spin-symmetry in the spin-incoherent regime. 

The rest of the paper is organized as follows. In Sec. II, we introduce the single-particle density matrix for 1D SU($\kappa$) fermions in terms of separate spatial and spin parts of the density matrix with anyonic statistics \cite{Yang2017, Marmorini2016, Hao2016}. In Sec. III, we investigate two cases for the spin parts of the density matrix, which are, respectively, the case of equal populations for each spin components and the other one involving all $S_z$ manifolds. We then show the numerically calculated momentum distributions and high momentum tails in Sec. IV, and compare the tails with analytical predictions. Finally we conclude in Sec. V.  

\section{Single-particle density matrix of SILL SU(\texorpdfstring{$\boldsymbol{\kappa}$}{kappa}) fermions}

The effective Hamiltonian of ultracold 1D SU($\kappa$) fermions in TG gas limit can be expressed as \cite{Pagano2014, Decamp2016},
\bea
H&=&\sum_{\nu=1}^\kappa\sum_{j=1}^{N_\nu}\left[-\frac{\hbar^2}{2m}\frac{\partial^2}{\partial x_{j,\nu}^2}+\frac{1}{2}m\omega^2x_{j,\nu}^2\right]\mathbb{I}_{\rm spin}\nonumber\\
&+&\sum_{\nu<\nu'}^\kappa\sum_{j=1}^{N_\nu}\sum_{j'=1}^{N_{\nu'}}\delta(x_{j,\nu}-x_{j',\nu'})g_{1D}\mathbb{I}_{\rm spin},
\eea
where we consider the atoms, with mass $m$, trapped in a harmonic potential with the axial trap frequency $\omega$, and $\kappa$ spin components satisfy $\sum_{\nu=1}^\kappa N_\nu=N$ with the number of atoms $N_\nu$ for $\nu$th component. The spin-independent interactions between SU($\kappa$) spin-symmetric fermions can be described by $g_{1D}=-2\hbar^2/(ma_{1D})$, where $a_{1D}$ is the effective scattering length \cite{Olshanii1998} in 1D. Next we consider a general wave function of $N$ fermions with spins,
\bea
|\Psi\rangle=\sum_{s_1,s_2,...s_N}\psi_{s_1,s_2,...s_N}(\vec{x})|s_1,s_2,...,s_N\rangle,\label{wf}
\eea
where we denote the atomic spatial distributions as $\vec x=(x_1,x_2,...,x_N)$ along with corresponding spin configurations $|s_1,s_2,...,s_N\rangle\equiv|\vec s\rangle$. Note that each spin $s_i$ is within the manifold of SU($\kappa$) spin symmetry. The total wave function must satisfy the quantum statistics of the atoms, which is fermionic anti-symmetry considered here, and thus it is sufficient if we only focus on the ordered region of $x_1$ $<$ $x_2$ $<$ $...$ $<$ $x_N$. The other regions can be obtained via permutations of this ordered region.  

The single-particle density matrix according to the general wave function of Eq. (\ref{wf}) becomes 
\bea
\rho(x',x)=N\sum_{\vec{s}}\int d\bar{x}\psi_{\vec{s}}^*(x',\bar{x})\psi_{\vec{s}}(x,\bar{x}),\label{rhop}
\eea 
where $\bar{x}$ $\equiv$ $(x_2,x_3,...,x_N)$. To proceed to calculate Eq. (\ref{rhop}), we consider only the region of $x'<x$ which is symmetric to $x'>x$. Equation \ref{rhop} involves $N(N+1)/2$ distinct and ordered integral regions \cite{Jen2016_spin1, Jen2017_spin1}, which we denote as \cite{Yang2017} 
\bea
\Gamma_{m,n}:~&&x_2<...<x_m<x'<x_{m+1}<...\nonumber\\
             &&...<x_n<x<x_{n+1}...<x_N,
\eea 
where $x'$ and $x$ are located right behind $x_m$ and $x_n$ respectively. Each distinct and ordered integral region has the same spatial integral value, and such that we obtain \cite{Jen2016_spin1, Jen2017_spin1, Yang2017}
\bea
\rho(x'<x)=\sum_{m=1}^N\sum_{n=m}^N\rho_{m,n}(x',x)S_{m,n}.\label{rho2}
\eea
In the above, we proceed to write down the spatial part in TG gas limit as
\bea
\rho_{m,n}(x',x)=&&(-1)^{n-m}N!\int_{\Gamma_{m,n}}d\bar x\varphi_{\vec n}^*(x',\bar x)\varphi_{\vec n}(x,\bar x),\label{rhomn}\\
\varphi_{\vec n}(\vec x)\equiv&&\frac{1}{\sqrt{N!}}\mathbb{A}[\phi_{n_1}(x_1),\phi_{n_2}(x_2),...,\phi_{n_N}(x_N)],\label{psi}
\eea
with orbital indices $\vec n$ $=$ $(n_1,n_2,...,n_N)$ and antisymmetrized ($\mathbb{A}$) eigenfunctions $\phi_{n_j}(x_j)$ of noninteracting fermions in a harmonic trap. Meanwhile, the spin part in SILL regime is denoted as \cite{Jen2016_spin1, Jen2017_spin1}
\bea
S_{m,n}=(-1)^{m-n}\frac{\sum_{\vec s}\langle P_{12...m}(\vec s)|P_{12...n}(\vec s)\rangle}{\rm{Tr}_\chi(E)},\label{Smn}
\eea
with identical and $m$-particle permutation operators $E$ and $P_{12...m}$ respectively. The total number of spin state configurations is Tr$_\chi(E)\equiv\sum_\chi\langle\chi|E|\chi\rangle$ for all spin configurations $|\chi\rangle$. The $S_{m,n}$ represents the normalized spin function overlaps, which is averaged by all possible spin configurations and is nonvanishing if the permuted spins $|P_{12...m}(\vec s)\rangle$ has projections on $|P_{12...n}(\vec s)\rangle$.

To evaluate Eq. (\ref{rho2}) efficiently, we take advantage of the discrete Fourier transform or equivalently anyonic statistics \cite{Yang2017, Marmorini2016, Hao2016}, which transforms respectively Eqs. (\ref{rhomn}) and (\ref{Smn}) to 
\bea
\rho_{m,n}(x',x)=&&N^{-2}\sum_{k',k}\rho_{k',k}(x',x)e^{i\pi k' m}e^{-i\pi k n},\\
S_{k',k}=&&N^{-2}\sum_{m,n=1}^N S_{m,n}e^{i\pi k' m}e^{-i\pi k n},
\eea 
with discrete statistical parameters \cite{Girardeau2006} of $k,k'=2j/N$ for $j=1,2,...,N$, and 
\bea
\rho_{k',k}(x',x)=&&N\int d\bar x\prod_{j=2}^N A^{k'*}(x_j-x')A^{k}(x_j-x)\nonumber\\
&&\times\varphi_{\vec n}^*(x',\bar x)\varphi_{\vec n}(x,\bar x),\label{rhokk}
\eea
where $A^{k}(x_j-x_l)\equiv e^{i\pi(1-k)\theta(x_j-x_l)}$ with the Heaviside step function $\theta(x_j-x_l)$. Finally, we obtain the single-particle density matrix for SILL 1D SU($\kappa$) fermions, in terms of discrete statistical parameters, 
\bea
\rho(x',x)=\sum_{k',k}\rho_{k',k}(x',x)S_{k',k}.\label{rhoxx}
\eea 
In the next section, we specifically calculate the spatial and spin parts of the density matrix for SU($\kappa$) fermions.

\section{Spatial and Spin parts of the density matrix}

\subsection{Spatial parts of the density matrix}

The spatial parts of the single-particle density matrix have been investigated for spinless bosons \cite{Papenbrock2003, Forrester2003} and anyons \cite{Hao2016,Marmorini2016} in a harmonic trap, where analytically exact formulas can be derived. For 1D SU($\kappa$) fermions in the TG gas limit and confined in a harmonic trap potential, the dimensionless eigenfunctions $\phi_n(y)$ with $y\equiv x/x_{ho}$ and $x_{ho}\equiv\sqrt{\hbar/(m\omega)}$ are
\bea
\phi_n(y)&=&\frac{1}{\sqrt{2^n n!}}\frac{1}{\pi^{1/4}}H_n(y)e^{-y^2/2},\label{eigen}
\eea
where $H_n$ are Hermite polynomials.

Put Eq. (\ref{eigen}) into Eq. (\ref{psi}), the spatial wave function with $\vec n =(0,1,...,N-1)$ can be expressed in a form of Vandermonde determinant \cite{Forrester2003},
\bea
\varphi_{\vec n}(\vec x)=\sqrt{C_N^V}\prod_{l=1}^Ne^{-x_l^2/2}\prod_{1\leq j<m\leq N}(x_j-x_m),\label{V}
\eea
where the normalization constant is
\bea
C_N^V=\frac{2^{N(N-1)/2}}{\pi^{N/2}\prod_{j=1}^N j!}.
\eea
To derive the exact form of Eq. (\ref{rhokk}), in addition to using the above form, we need the following general equality \cite{Papenbrock2003, Forrester2003},
\bea
&&\frac{1}{N!}\prod_{l=1}^N\int_{-\infty}^\infty dx_lg(x_l)(\textrm{det}[f_{j-1}(x_m)]_{j,m=1,...,N})^2\nonumber\\
&&=\textrm{det}\left[\int_{-\infty}^\infty dt g(t)f_{j-1}(t)f_{m-1}(t)\right]_{j,m=1,...,N},\label{equal}
\eea
for any functions $g$ and $f_j$. We separate the dependence of $x$ and $\bar x$ in $\varphi_{\vec n}(x,\bar x)$ by interpreting it in terms of minors, $\phi_n(x)\textrm{det}[\phi_{j}(x_m)]$ with $j=0,...,n-1,n+1,...,N-1$ and $m=2,...,N$, respectively for $n=0,1,...,N-1$. This way we are able to cast $\phi_n(x)$ in a Vandermonde form, while retain the rest of particles at $\bar x$ in a determinant form. Similar treatment to $\varphi_{\vec n}(x',\bar x)$ can be done. We then re-express Eq. (\ref{rhokk}) by grouping anyonic statistics of $A^{k'*}$ and $A^{k}$ with $(x_l-x)$ and $(x_l-x')$, and let $g(x_l)=A^{k'*}(x_l-x')A^{k}(x_l-x)(x_l-x')(x_l-x)$ in Eq. (\ref{equal}) with $l$ starting from $2$. Applying the equality of Eq. (\ref{equal}) to Eq. (\ref{rhokk}), we obtain
\begin{widetext}
\bea
\rho_{k',k}(x',x)=&&\frac{2^{N-1}}{(N-1)!}\frac{e^{-(x'^2+x^2)/2}}{\sqrt{\pi}}\textrm{det}\left[\int_{-\infty}^\infty dtA^{k'*}(t-x')A^{k}(t-x)(t-x')(t-x)\phi_j^*(t)\phi_m(t)\right]_{j,m=0,...,N-2},\nonumber\\
&&=\frac{e^{-(x'^2+x^2)/2}}{\sqrt{\pi}}\textrm{det}\left[\frac{2^{(j+m)/2}}{\Gamma(j+1)\Gamma(m+1)}\frac{b_{j,m}^{k',k}(x',x)}{\sqrt{\pi}}\right]_{j,m=1,...,N-1},
\eea
\end{widetext}
where
\bea
b_{j,m}^{k',k}(x',x)\equiv&&\int_{-\infty}^\infty dt A^{k'*}(t-x')A^{k}(t-x)\nonumber\\
&&\times(t-x')(t-x)t^{j+m-2}e^{-t^2}.\label{b}
\eea
Equation (\ref{b}) can be derived by using equivalent Vandermonde determinant forms for det[$H_{j-1}(x_m)$] and det[$x_m^{j-1}$] which leads to the term of $t^{j+m-2}$. We further derive the exact form of $b_{j,m}^{k',k}(x',x)$ in Appendix, which involves special functions of incomplete gamma functions. This exact form significantly expedites the numerical calculations of single-particle density matrix, but as $N$ increases and larger than $N=32$, the calculation is limited by $64$-digit computer double-precision. And as such, we give results up to $N=32$, which however can be pushed further by using arbitrary precision protocols. Next we study two cases of the spin parts in the density matrix of 1D SU($\kappa$) fermions.
 
\subsection{Equal populations in each spin components}

For equal populations in each SU($\kappa$) spin components, the spin configurations which contribute to Eq. (\ref{Smn}) involve the states,
\bea
|\underbrace{\alpha_1...\alpha_1}_{N_1}\underbrace{\alpha_2...\alpha_2}_{N_1}...\underbrace{\alpha_\kappa...\alpha_\kappa}_{N_1}\rangle,
\eea
where $N_1=N/\kappa$ and $\kappa$ spin components $(\alpha_1,\alpha_2,...,\alpha_\kappa)$. The $S_{m,n}$ is nonvanishing only when $N_1\geq l$ with $l\equiv|n-m|+1$ from the contributions of $l$ entries in each spin components. Take component $\alpha_1$ as an example, the contributing spin configuration is
\bea
|\underbrace{\alpha_1...\alpha_1}_{l}\underbrace{\alpha_1...\alpha_1}_{N_1-l}\underbrace{\alpha_2...\alpha_2}_{N_1}...\underbrace{\alpha_\kappa...\alpha_\kappa}_{N_1}\rangle.
\eea
Since there are $\kappa$ spin components, we obtain the spin parts of the density matrix as
\bea
S_{m,n}=\frac{(-1)^{m-n}}{w_N}\left[\frac{\kappa(N-l)!}{(N_1-l)![N_1!]^{\kappa-1}}\right],\label{S}
\eea 
where
\bea
w_N=\frac{N!}{[N_1!]^\kappa}.
\eea 
The bracket in Eq. (\ref{S}) originates from the number of states obtained by permuting the rest of ($N_1-l$) spins for one of the spin component and other $N_1$ spins from $(\kappa-1)$ components. We also define $w_N\equiv\textrm{Tr}_\chi(E)$ as the total number of states in the above. 

\subsection{All \texorpdfstring{$\boldsymbol{S_z}$}{Sz} manifolds included}

Next we consider the spin configurations with all $S_z$ manifolds. In contrast to the case of equal populations, the $S_{m,n}$ here is always finite, which has a contribution from $l\equiv|n-m|+1$ entries in each spin components. Take again the component $\alpha_1$ as an example, the contributing spin configuration is
\bea
|\underbrace{\alpha_1...\alpha_1}_{l}\underbrace{......}_{N-l}\rangle,
\eea
where the rest of $(N-l)$ spin components can be any of $\kappa$ ones. We obtain the spin parts of the density matrix as
\bea
S_{m,n}=&&\frac{(-1)^{m-n}}{w_N}\kappa^{N-l+1},\label{S2}\\
w_N=&&\kappa^N,
\eea 
and we can further simply $S_{m,n}$ as
\bea
S_{m,n}=&&\frac{(-1)^{m-n}}{\kappa^{|m-n|}}.\label{S3}
\eea 

\section{Momentum distributions}

Based on Eq. (\ref{rhoxx}), we numerically calculate the momentum distributions of SILL 1D SU($\kappa$) fermions in TG gas limit ($\hbar$ $=$ $1$), 
\bea
\rho(p)=\frac{1}{2\pi}\int_{-\infty}^\infty dx' \int_{-\infty}^\infty dx e^{ip(x'-x)}\rho(x',x).
\eea
Below we investigate various conditions of fixed total number of atoms $N$, spin components $\kappa$, and number of atoms in each spin components $N_1$. 

\subsection{Fixed \texorpdfstring{$\boldsymbol{N}$}{N}}

\begin{figure}[t]
\centering
\includegraphics[width=8.5cm,height=4.5cm]{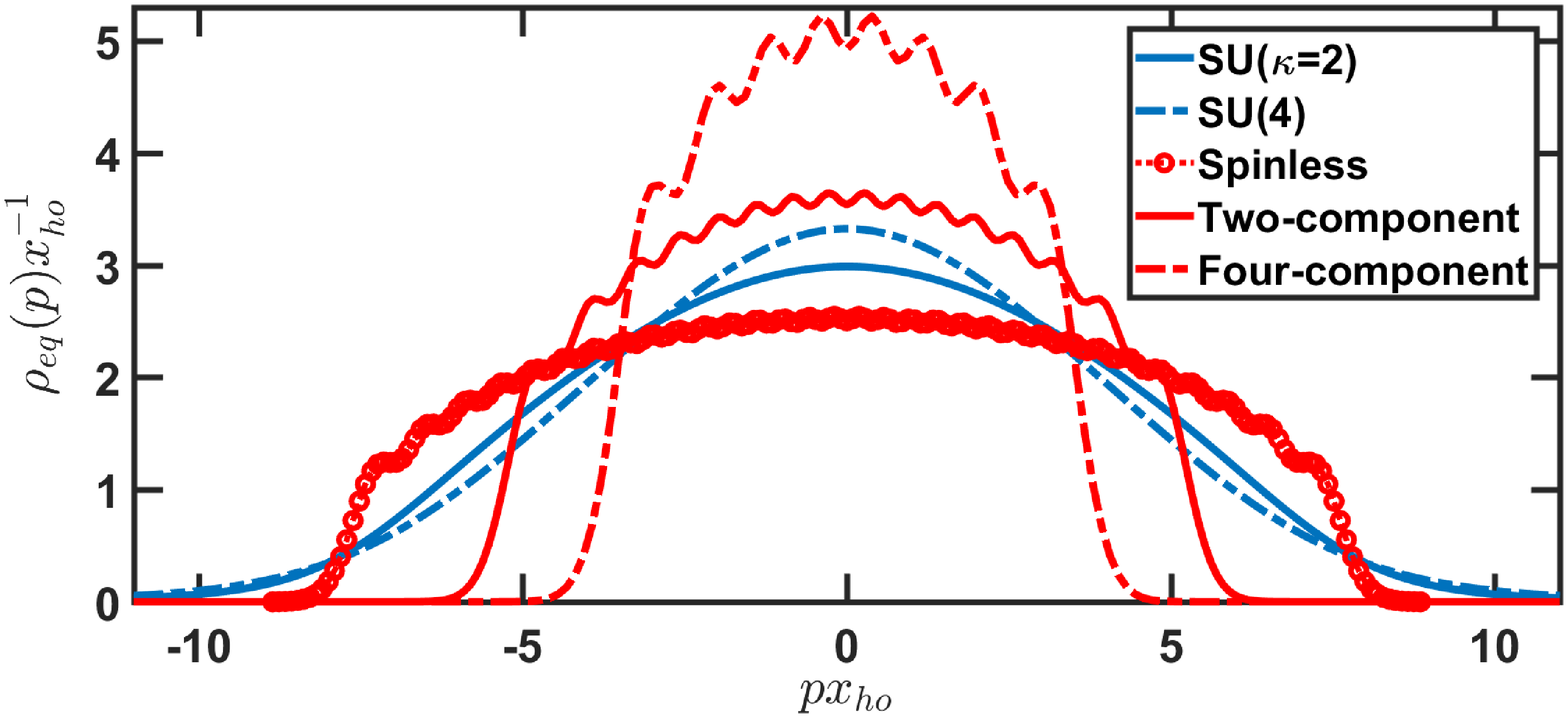}
\caption{Comparison of momentum distributions of 1D SU($\kappa$) fermions in the SILL regime and noninteracting multi-component fermions. The number of fermions is $N=32$. As the number of components increases, the width of momentum distributions decreases for both noninteracting and spin-incoherent cases.}\label{fig1}
\end{figure}

It is instructive at first to compare the momentum distributions of 1D SU($\kappa$) fermions in TG gas limit with noninteracting multi-component ones. In Fig. \ref{fig1}, we focus on the case of equal populations in each spin components [$\rho_{eq}(p)$]. The feature of noninteracting $\kappa$-component fermions manifests in the number of Friedel oscillation peaks, which is exactly $N_1$ of them. As the number of components increases, noninteracting fermions tend to occupy lower momenta, and thus the width of momentum distributions becomes narrower. This can be explained by the decreasing $N_1$ for each spin components with a fixed $N$. This trend is also seen in SILL 1D SU($\kappa$) fermions as $\kappa$ increases. In contrast, for fixed $\kappa$ component and $N$, TG gas has a broadened width of momentum distributions compared to the ones of noninteracting fermions due to the strong interactions in TG gas limit. Furthermore, we note that the kinetic and potential energies of 1D SU$(\kappa)$ fermions in TG gas limit satisfy the virial theorem \cite{Werner2006, Werner2008}, which are equally $N^2\hbar \omega/4$ (half of the total energy of the system) since the fermions have the same density profile as the noninteracting ones. Instead for noninteracting multi-component fermions, the kinetic (potential) energy is $N^2\hbar\omega/(4\kappa)$ which is always smaller than the one of SILL SU($\kappa$) fermions for $\kappa\geq 2$.

\begin{figure}[t]
\centering
\includegraphics[width=8.5cm,height=4.5cm]{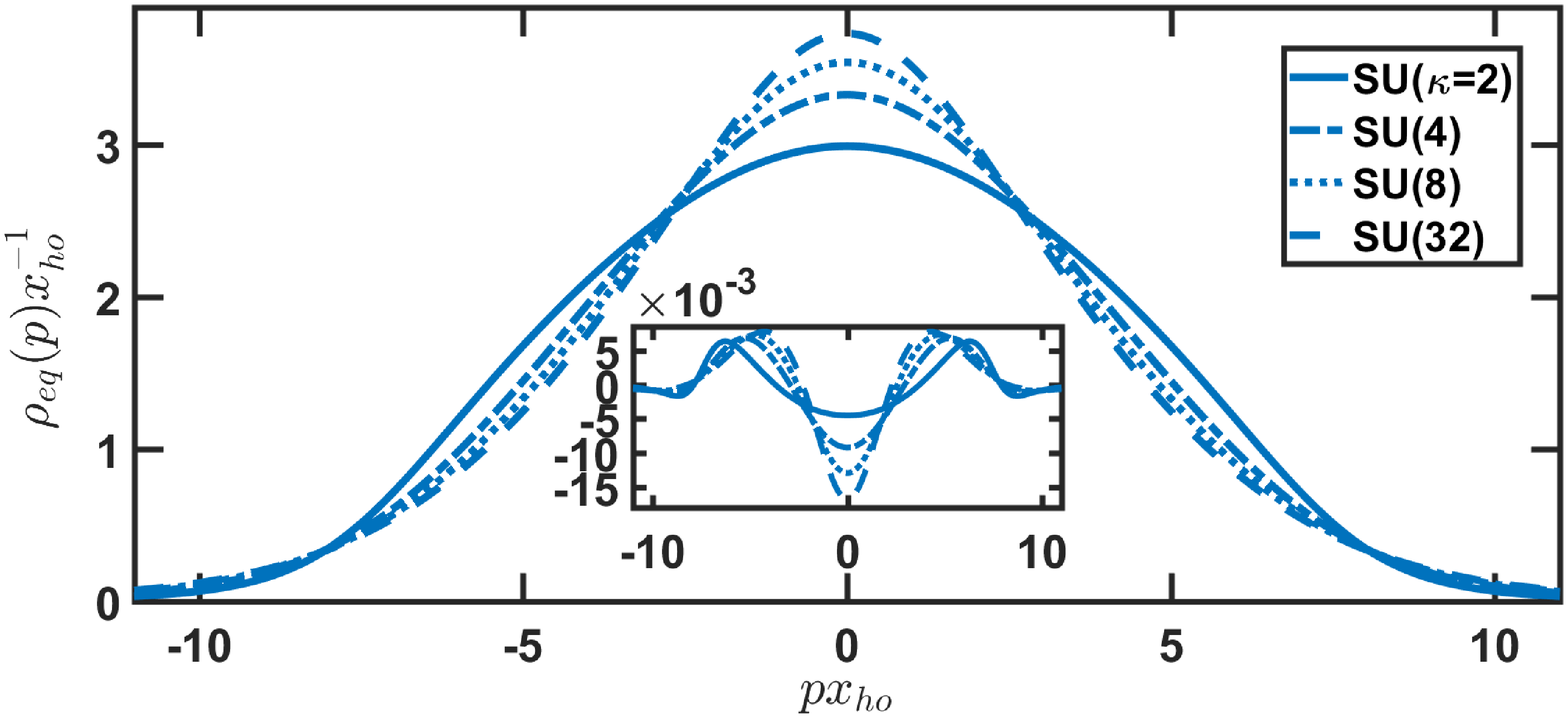}
\caption{Momentum distributions of 1D SU($\kappa$) fermions in the SILL regime with equal populations in each spin components. The number of fermions is the same as Fig. \ref{fig1}, and we choose $\kappa=2,4,8,32$ to compare with spinless fermions. The inset shows the difference between the cases of all spin manifolds and equal populations, $[\rho_{all}(p)-\rho_{eq}(p)]/\rho_{eq}(0)$, which is order of $10^{-3}$ to the maximum of the distribution, respectively.}\label{fig2}
\end{figure}

In Fig. \ref{fig2}, we show $\rho_{eq}(p)$ with the same $N$ for different number of spin components. In contrast to the Friedel oscillations of spinless fermions, the oscillations in SILL SU($\kappa$) fermions are smoothed out due to the averaging effect of spin function overlaps $S_{m,n}$, similar to the case of spin-$1$ bosons \cite{Jen2016_spin1, Jen2017_spin1}. As $\kappa$ increases, the momentum distributions are less broadened, which can be seen near $px_{ho}\approx 5$ and also reflects on increasing $\rho_{eq}(p=0)$. In contrast, the high momentum tails have larger values for larger $\kappa$, which we will investigate further in details in the next subsection. For the case of all $S_z$ manifolds included, $\rho_{all}(p)$, we show its difference from $\rho_{eq}(p)$ in the insets of Fig. \ref{fig2}. The ratio of relative difference to $\rho_{eq}(0)$ is on the order of $10^{-3}$, which therefore makes $\rho_{all}(p)$ almost indistinguishable from $\rho_{eq}(p)$. Nonetheless, central maximum of $\rho_{all}(0)$ is smaller than $\rho_{eq}(0)$ in respective $\kappa$ components, while at moderate $3 \leq px_{ho}\leq 8$, $\rho_{all}(p)$ becomes larger than $\rho_{eq}(p)$ between around the two crossing points.

As a theoretical interest, we consider the case with a large $\kappa$. In Fig. \ref{fig2}, we show the momentum distribution with $\kappa$ $=$ $N$. For equal populations and fixed $N$, this case represents the maximal $\kappa$ allowed and indicates that every fermion occupies exactly one distinct spin. And as such, $\rho_{eq,\kappa=N}(p)$ has the narrowest width compared to all other $\kappa$ $<$ $N$. Again, $\rho_{all,\kappa=N}(p)$ does not distinguish much from $\rho_{eq,\kappa=N}(p)$ as shown in the inset of Fig. \ref{fig2}. At $\kappa=N$, for equal populations, we have $S_{m,n}=\delta_{m,n}$. Comparing this with Eq. (\ref{S3}) shows that $\rho_{all, \kappa\rightarrow\infty}(x',x)$ coincides exactly with $\rho_{eq,\kappa=N}(x',x)$, and so almost is indistinguishable from $\rho_{all,\kappa=N}(x',x)$. These narrower widths and higher momentum tails are reminiscent of the infinite $\kappa$ regime where the ground state energy \cite{Yang2011} and Tan's contacts \cite{Decamp2016} of 1D SU($\kappa$) fermions approach the case of spinless bosons. However, for SILL 1D SU($\kappa$) fermions at infinite $\kappa$, the spin parts of the density matrix for all spin manifolds become $S_{m,n}\rightarrow\delta_{m,n}$, whereas for spinless bosons, $S_{m,n}$ $=$ $1$ for all $m$ and $n$. Therefore, SILL 1D SU($\kappa$) fermions never behave exactly as spinless bosons as $\kappa\rightarrow\infty$. Under this limit, we note that $S_{k',k}\rightarrow\delta_{k',k}/N$, and single particle matrix becomes $\rho_{all,\kappa\rightarrow\infty}(x',x)=N^{-1}\sum_{k}\rho_{k,k}(x',x)$, an average over anyon density matrices with statistical parameters $k$. 

\subsection{Fixed \texorpdfstring{$\boldsymbol{\kappa}$}{kappa}}

\begin{figure}[t]
\centering
\includegraphics[width=8.5cm,height=4.5cm]{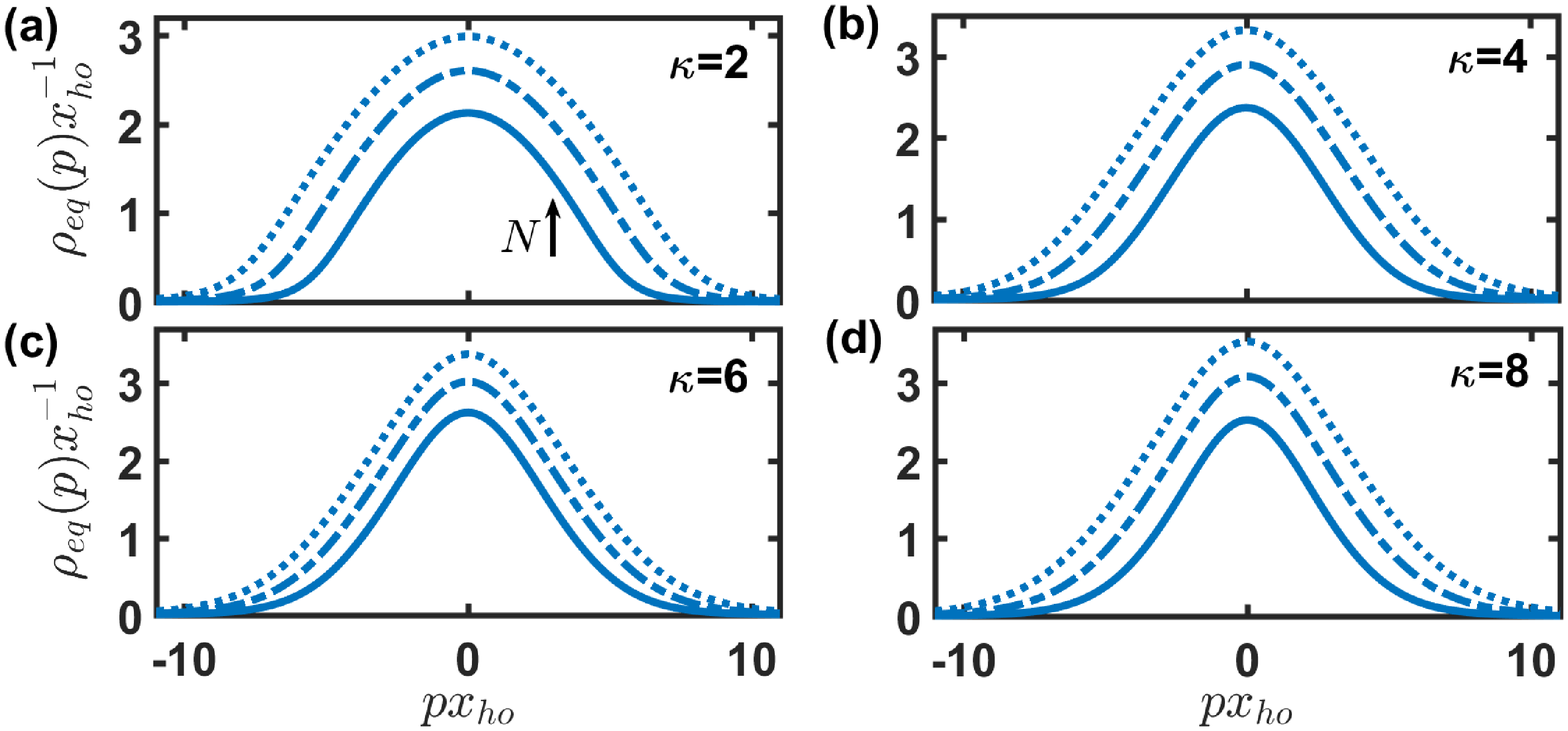}
\caption{Momentum distributions of SILL 1D SU($\kappa$) fermions for various $N$ with equal populations in each spin components. As the number of fermions $N$ increases, the momentum distributions are uniformly broadened. For various spin components $\kappa$, we choose $N=16,24,32$ in (a), (b), (d), and $18,24,30$ in (c), respectively.}\label{fig3}
\end{figure}

In Fig. \ref{fig3}, we show $\rho_{eq}(p)$ with fixed number of spin components $\kappa$. They are broadened uniformly as $N$ increases for various spin components due to strong interactions. We know that for noninteracting multi-component fermions, their peaks scale as $N^{0.5}$, while by fitting our numerical results in Fig. \ref{fig3}, we find that the peaks scale as $N^\alpha$ where $\alpha \lesssim 0.5$. 

\subsection{Fixed \texorpdfstring{$\boldsymbol{N_1}$}{N1}}

\begin{figure}[b]
\centering
\includegraphics[width=8.5cm,height=4.5cm]{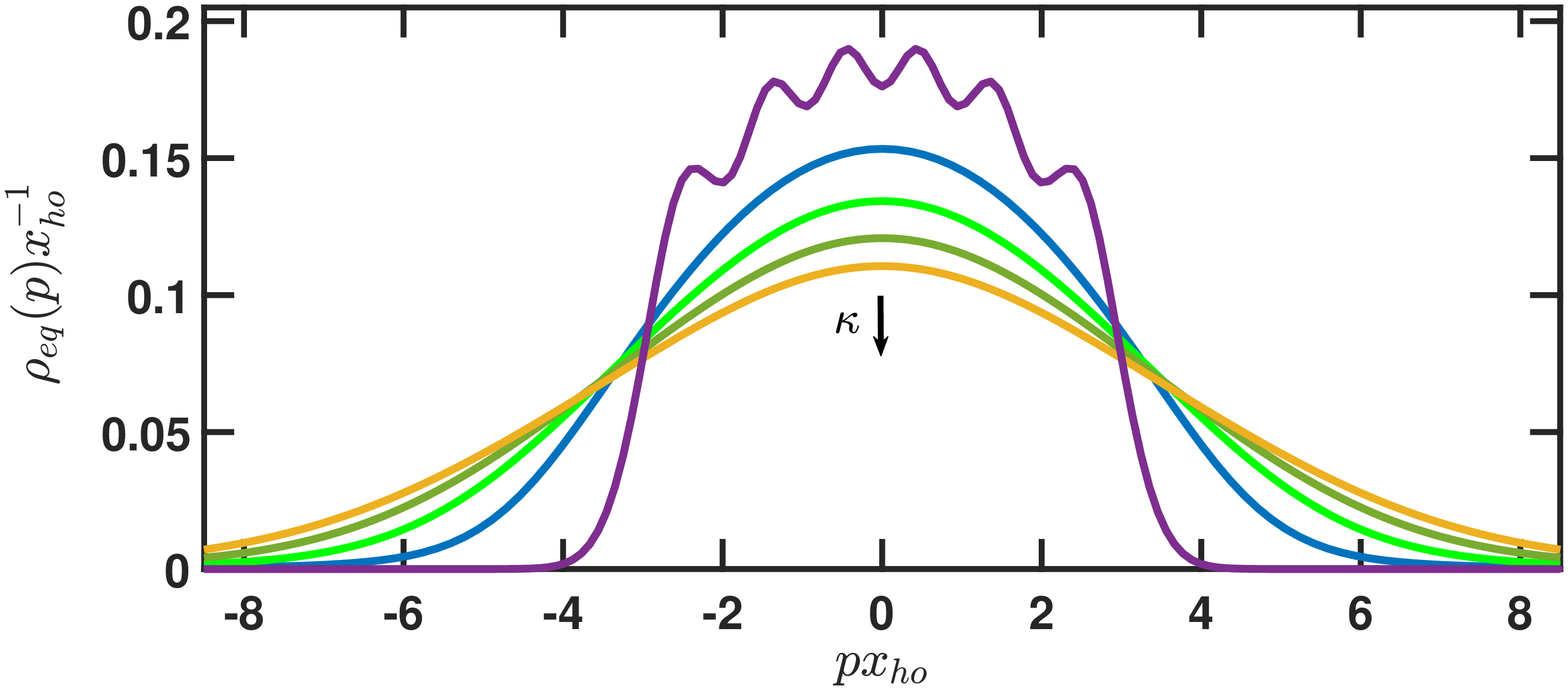}
\caption{Normalized momentum distributions of 1D SU($\kappa$) fermions in the SILL regime with fixed number of atoms per component. The number of atoms per spin component is $N_1=6$. As the number of components $\kappa$ increases from $1$ to $5$ and accordingly $N=\kappa N_1$, the momentum distributions are broadened.}\label{fig4}
\end{figure}

Finally, as in the experiment of 1D fermions with tunable SU($\kappa$) spin symmetry \cite{Pagano2014}, in Fig. \ref{fig4} we plot the normalized momentum distributions ($\int \rho(p)dp=1$) of SILL 1D SU($\kappa$) fermions with fixed $N_1$. We see broadening in momentum distributions as $\kappa$ increases. As spin components of the fermions increase, the total number of atoms also increase. Therefore, the broadening of momentum distributions comes both from strong interactions of fermions in TG gas limit and increasing number of atoms. Under the condition of a fixed $N_1$ in Fig. \ref{fig4}, the kinetic (potential) energy per atom is $N_1\kappa\hbar\omega/4$, which rises up linearly as $\kappa$ increases. 

We also compare our results with the experiments, where the system is at finite temperature with finite atom-atom interactions, and experiences inhomogeneous distributions in 2D optical lattice of 1D tubes \cite{Pagano2014}. In the experiment, the broadening of the normalized momentum distributions is also observed as $\kappa$ increases, though the Friedel oscillation is absent in single component measurement due to the averaging of inhomogeneous distributions or finite temperature. In addition, we extrapolate their normalized momentum distributions in Fig. 2(a) of \cite{Pagano2014}, and numerically calculate their kinetic energies. The kinetic energies approximately follow the linear increase of $\kappa$ spin components, which indicates that the behavior of the system is similar as in TG gas limit. We note that the other essential feature of SILL 1D SU($\kappa$) fermions manifests in the trend of momentum distributions toward narrower ones for a fixed $N$ in Fig. \ref{fig2}, as an alternative method to measuring breathing mode oscillations \cite{Pagano2014}. 

\subsection{Large \texorpdfstring{$\boldsymbol{p}$}{p} asymptotics}

Here we further investigate the high momentum tails of 1D SU($\kappa$) fermions in the SILL regime. This universal high momentum asymptotic $1/p^4$ originates from many-body systems with two-body contact interaction, which is present in a spinless Bose gas \cite{Minguzzi2002, Olshanii2003, Xu2015, Lang2017, Decamp2018}, SILL spin-$1$ Bose gas \cite{Jen2016_spin1, Jen2017_spin1}, two-component \cite{Braaten2008-1, Braaten2008-2,Werner2009, Zhang2009, Patu2016} or multi-component Fermi gas \cite{Matveeva2016, Decamp2016}, and Tan's relation \cite{Tan2008, Barth2011}. The coefficients of the scaling can be related to the slope of the ground state energy of the many-body system, that is $(-dE/dg_{\rm 1D}^{-1})$ \cite{Olshanii2003, Decamp2016}.

We have derived the analytical results for this high momentum asymptotic in SILL 1D spin-$1$ TG Bose gas \cite{Jen2016_spin1, Jen2017_spin1}, which can be straightforwardly extended to 1D SU($\kappa$) fermions in TG gas limit,
\bea
\rho(p)\underset{p\rightarrow\infty}{=}&&\frac{2(1+S_{m-1,m})}{2\pi p^4}\nonumber\\
&&\times\sum_{(n_i,n_j)}\int_{-\infty}^\infty dx
\left| \begin{array}{cc}
\phi_{n_i}'(x) & \phi_{n_j}'(x) \\
\phi_{n_i}(x) & \phi_{n_j}(x) \end{array} \right|^2,\label{large_p}
\eea 
for arbitrary $m$ since $S_{m,n}$ only depends on $|m-n|$. The $(n_i,n_j)$ represents all possible pairs of $N$ harmonic oscillator eigenfunctions. The coefficients depend only on the spin parts of the single-particle density matrix, $S_{m,n}$, with $|m-n|=1$, since they have the contributions only from the integral regions of $x$ $<$ $x_j$ $<$ $x'$ and $x'$ $<$ $x_j$ $<$ $x$ for all $x_j$ $\in$ $\bar{x}$ with $x$ $\approx$ $x'$. The coefficients for spinless or spinful bosons can be obtained by replacing $(1+S_{m-1,m})$ with $2$ or by $S_{m-1,m}\rightarrow -S_{m-1,m}$ respectively in Eq. (\ref{large_p}). The sign change of $S_{m-1,m}$ for spinful bosons restores the bosonic symmetry in the single-particle density matrix. For fermions, we note that the coefficients of $(1+S_{m-1,m})$ in the asymptotic forms of Eq. (\ref{large_p}) increase as $\kappa$ increases. At $\kappa\rightarrow\infty$, $S_{m-1,m}\rightarrow 0$, and such that $(1+S_{m-1,m})$ maximizes to be one but is only half of the coefficient for spinless bosons. This value is thus also half of that of the ground state of 1D fermions in the $\kappa\rightarrow\infty$, TG limit.

\begin{figure}[t]
\centering
\includegraphics[width=8.5cm,height=4.5cm]{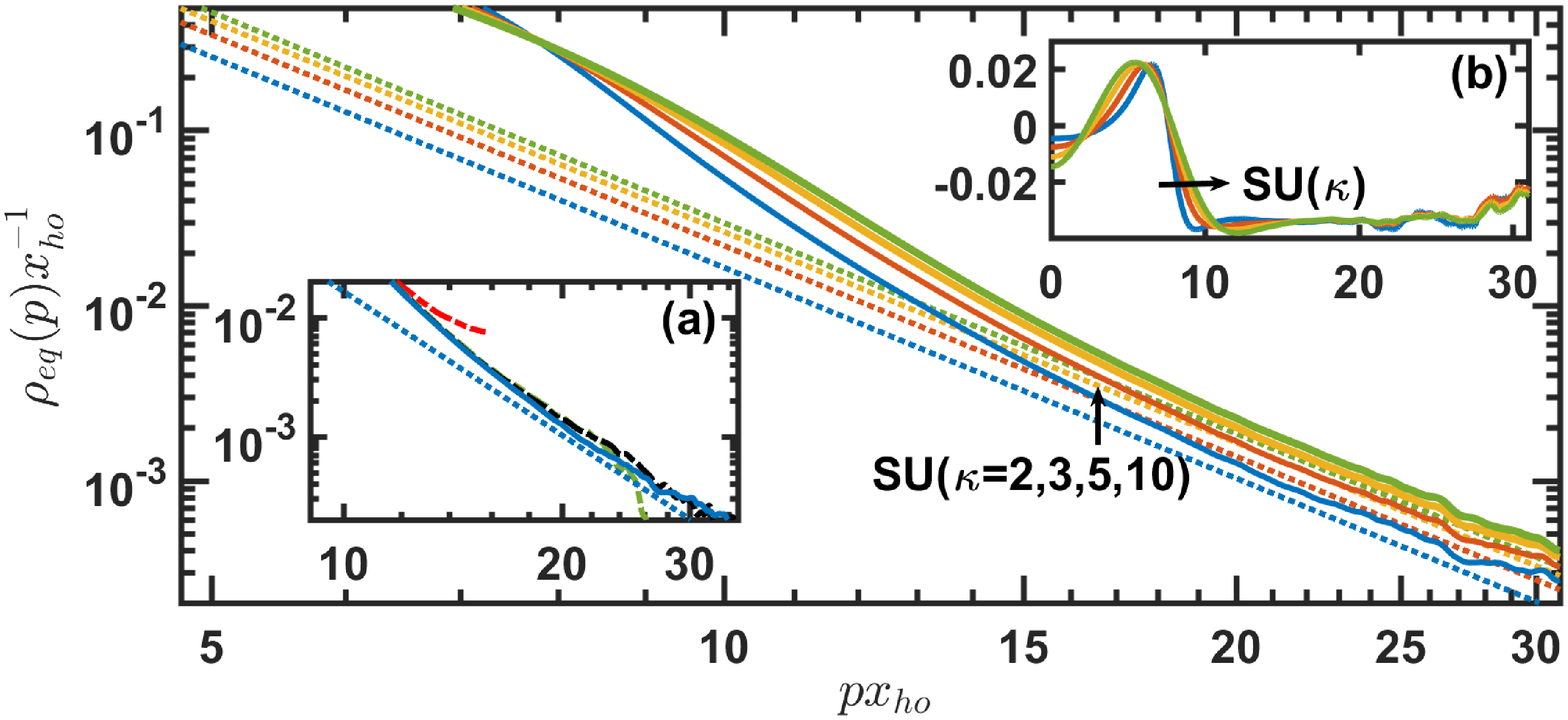}
\caption{Asymptotics of high momentum distributions of 1D SU($\kappa$) fermions in the SILL regime. The total number of fermions is $N=30$, and we choose $\kappa=2,3,5,10$ for comparisons. High momentum tails are plotted in logarithmic scales compared with analytical curves (dash). The analytical asymptotics for $\rho_{eq}(\infty)$ are $(165, 220, 264, 297)/p^4$ respectively as $\kappa$ increases, while for $\rho_{all}(\infty)$, they are $(159, 213, 255, 287)/p^4$. The inset (a) shows the convergence of the numerical result of SU($\kappa=2$) to the analytical curve as finer grids increases from $dx=0.2,0.1,0.05$ (dash dots) to $0.025$ (solid), while (b) shows the relative asymptotics of $[\rho_{all}(p)-\rho_{eq}(p)]/\rho_{eq}(p)$.}\label{fig5}
\end{figure}

In Fig. \ref{fig5}, we show the high momentum asymptotic curves and compare them with the analytical results. As $\kappa$ increases, the coefficients go up as the arrow indicates. The convergence of the numerical calculations can be seen in the inset (a) where the numerical result approaches the analytical one as finer grids are used. We further compare $\rho_{all}(p)$ with $\rho_{eq}(p)$ by showing its relative difference in the inset (b). The coefficients for high momentum tails of $\rho_{all}(p)$ are smaller than the case of $\rho_{eq}(p)$. Similar to the momentum distributions at small $p\lesssim 10$, the difference ratio is of order of $10^{-2}$ relative to $\rho_{eq}(p)$ at large $p$, and therefore the asymptotic curves of $\rho_{all}(p)$ is again close to the ones of $\rho_{eq}(p)$. The analytical results of the coefficients also show only a relative difference of less than $5\%$ (see caption of Fig. \ref{fig5} for numerical vales of the coefficients), which almost overlap with each other. At $10 \lesssim px_{ho}\lesssim 25$, the relative difference of inset (b) saturates to a flat line, indicating the constant ratio of the coefficients between $\rho_{all}(p)$ and $\rho_{eq}(p)$. Meanwhile, for $px_{ho}\gtrsim 25$, this difference goes up, which marks the accuracy range of $px_{ho}\approx 25$ in our numerical calculations.

\section{Conclusion}

In conclusion, we have investigated the momentum distributions of 1D SU($\kappa$) fermions in TG gas limit, which puts the system in a spin incoherent regime, forming a different universal class of SILL from conventional Luttinger liquid. We derive the single-particle density matrices in terms of those of anyons, which help expedite the numerical calculations up to $N=32$. We further investigate SU($\kappa$) fermions in two cases of equal populations in each spin components and all $S_z$ manifolds included. Compared to noninteracting multi-component fermions, their momentum distributions are broadened due to strong interactions in TG gas limit, while become less broadened as $\kappa$ increases. We also compare the numerical results with the analytical predictions in high momentum tails, which follow asymptotically the analytical coefficients we derived in moderately high momentum regions. Our results provide an informative comparison with experiments of multicomponent alkaline-earth fermions with SU($\kappa$) spin-symmetry in the spin-incoherent regime.

\section*{Acknowledgements}
This work is supported by the Ministry of Science and Technology (MOST), Taiwan, under Grant No. MOST-104-2112-M-001-006-MY3 and MOST-106-2112-M-001-005-MY3. H.H.J is partially supported by the Grant No. of MOST-106-2633-M-001-001 and 106-2811-M-001-130 from MOST, as an assistant research scholar in Institute of Physics, Academia Sinica, Taiwan.
\appendix*
\section{Exact form of \texorpdfstring{$\boldsymbol{b_{j,m}^{k',k}(x',x)}$}{b}}

We here derive the exact form of $b_{j,m}^{k',k}(x',x)$ of Eq. (\ref{b}) in the main text. Replacing $A^{k}(t-x)$ with $e^{i\pi(1-k)\theta(t-x)}$ and considering $x'<x$, we obtain
\bea
b_{j,m}^{k',k}(x',x)=&&\int_{-\infty}^\infty dt e^{-i\pi(1-k')\theta(t-x')}e^{i\pi(1-k)\theta(t-x)}\nonumber\\
&&\times(t-x')(t-x)t^{j+m-2}e^{-t^2},
\eea
which can be further decomposed into three integral regions,
\bea
b_{j,m}^{k',k}(x',x)=&&\left[\int_{-\infty}^\infty dt+(e^{-i\pi(1-k')}-1)\int_{x'}^x dt\right.\nonumber\\&&
\left.+(e^{i\pi(k'-k)}-1)\int_{x}^\infty dt\right](t-x')(t-x)\nonumber\\
&&\times t^{j+m-2}e^{-t^2}=A+B+C.\label{b2}
\eea

These integrals can be exactly expressed in terms of incomplete gamma functions. The definitions of upper/lower incomplete gamma functions and ordinary gamma function are defined respectively as
\bea
\Gamma(s,x)\equiv&&\int_x^\infty t^{s-1}e^{-t}dt ,\\
\gamma(s,x)\equiv&&\int_0^x t^{s-1}e^{-t}dt ,\\
\Gamma(s)\equiv&&\int_0^\infty t^{s-1}e^{-t}dt =\Gamma(s,x)+\gamma(s,x).
\eea
The first integral of Eq. (\ref{b2}) becomes
\bea
A=&&\frac{1}{4}\left[(e^{i(j+m)\pi}+1)(2xx'+j+m-1)\Gamma\left(\frac{j+m-1}{2}\right)\right.\nonumber\\
&&\left.+2(e^{i(j+m)\pi}-1)(x+x')\Gamma\left(\frac{j+m}{2}\right)\right],
\eea
where various ordinary gamma functions can be derived by change of variables $t^2\rightarrow t'$ in Eq. (\ref{b2}). The second integral becomes
\bea
B=&&\left(e^{-i\pi(1-k')}-1\right)\left[x'x\mu_{j+m-2}(x',x)+\mu_{j+m}(x',x)\right.\nonumber\\
&&\left.-(x'+x)\mu_{j+m-1}(x',x)\right],
\eea
with
\bea
\mu_m(x',x)\equiv&&\frac{\epsilon(x)^{m+1}}{2}\gamma\left(\frac{m+1}{2},x^2\right)\nonumber\\&&-\frac{\epsilon(x')^{m+1}}{2}\gamma\left(\frac{m+1}{2},x'^2\right),
\eea
where $\epsilon(x)$ is the sign function, and similarly, various lower incomplete gamma functions can be derived by change of variables. Finally the third integral of Eq. (\ref{b2}) becomes
\begin{widetext}
\bea
C=&&\frac{1}{2}\left\{\left[\Gamma\left(\frac{j+m+1}{2}\right)-\epsilon(x)^{j+m+1}\gamma\left(\frac{j+m+1}{2},x^2\right)\right]-(x'+x)\left[\Gamma\left(\frac{j+m}{2}\right)-\epsilon(x)^{j+m}\gamma\left(\frac{j+m}{2},x^2\right)\right]\right.\nonumber\\&&
\left.+x'x\left[\Gamma\left(\frac{j+m-1}{2}\right)-\epsilon(x)^{j+m-1}\gamma\left(\frac{j+m-1}{2},x^2\right)\right]\right\}.
\eea
\end{widetext}

\end{document}